\documentclass[conference,a4paper]{IEEEtran}
\IEEEoverridecommandlockouts

\usepackage{amsmath,amssymb,amsfonts}
\usepackage{algorithmic}
\usepackage{graphicx}
\usepackage{url}
\usepackage{tabularx}
\usepackage{multirow} 
\usepackage{makecell}
\usepackage{textcomp}
\usepackage{xcolor}
\def\BibTeX{{\rm B\kern-.05em{\sc i\kern-.025em b}\kern-.08em
    T\kern-.1667em\lower.7ex\hbox{E}\kern-.125emX}}
    
\newcommand{\fu}[1]{\textcolor{black}{#1}}
    
\begin{document}
\graphicspath{ {./image/} }

\title{Effective Anomaly Detection in Smart Home by Integrating Event Time Intervals\\

}

\author{\IEEEauthorblockN{Chenxu Jiang}
\IEEEauthorblockA{\textit{Dept. of Electrical and Computer Engineering} \\
\textit{Stevens Institute of Technology}\\
Hoboken, NJ, United States \\
cjiang22@stevens.edu}
\and
\IEEEauthorblockN{Chenglong Fu}
\IEEEauthorblockA{\textit{Dept. of Computer and Information Sciences} \\
\textit{Temple University}\\
Philadelphia, PA, United States \\
chenglong.fu@temple.edu}
\and
\IEEEauthorblockN{Zhenyu Zhao}
\IEEEauthorblockA{\textit{Dept. of Computer and Information Sciences} \\
\textit{Temple University}\\
Philadelphia, PA, United States \\
tuo36712@temple.edu}
\and
\IEEEauthorblockN{Xiaojiang Du}
\IEEEauthorblockA{\textit{Dept. of Electrical and Computer Engineering} \\
\textit{Stevens Institute of Technology}\\
Hoboken, NJ, United States \\
xdu16@stevens.edu}
\and
\IEEEauthorblockN{Yuede Ji}
\IEEEauthorblockA{\textit{Dept. of Computer Science and Engineering} \\
\textit{University of North Texas}\\
Denton, TX, United States \\
yuede.ji@unt.edu}
}

\maketitle

\begin{abstract} Smart home IoT systems and devices are susceptible to attacks and malfunctions. As a result, users' concerns about their security and safety issues arise along with the prevalence of smart home deployments. In a smart home, various anomalies (such as fire or flooding) could happen, due to cyber attacks, device malfunctions, or human mistakes. These concerns motivate researchers to propose various anomaly detection approaches.
\fu{Existing works on smart home anomaly detection focus on checking the sequence of IoT devices' events but leave out the temporal information of events. This limitation prevents them to detect anomalies that cause delay rather than missing/injecting  events.}
\fu{To fill this gap, } in this paper, we 
\fu{propose a novel anomaly detection method that takes the inter-event intervals into consideration. We propose an innovative metric to quantify the temporal similarity between two event sequences. We design a mechanism to learn the temporal patterns of event sequences of common daily activities. Delay-caused anomalies are detected by comparing the sequence with the learned patterns.}
\fu{We collect device events from a real-world testbed for training and testing. The experiment results show that our proposed method achieves accuracies of 93\%, 88\%, 89\% for three daily activities.}
\end{abstract}


\section{Introduction}
\fu{In recent years, smart home IoT systems become increasing popular in the consumer market. }\fu{In US, it is reported that smart home IoT devices have entered 43\% households in 2021 \cite{devpenet}. Conveniences brought by smart home platforms, like Alexa and Google Home, also incentive common users to automate their life with smart appliances and sensors.}

However, \fu{security and safety concerns also raise along with the prevalence of smart home IoT devices~\cite{fernandes2016security}. Due to limitations on cost and power supply, smart home IoT devices are long known to be unreliable and vulnerable to cyber-attacks, which allows attackers to impose threats to users' safety in the physical world~\cite{papernot2018sok}. Besides deliberate attacks, device malfunctions~\cite{hnat2011hitchhiker} happen even more frequently as some devices are working in harsh environment (e.g., moisture in bathrooms). These device attacks and faults, if not dealt timely, could cause severe damages. For example, with the help of automation rules~\cite{STPub}, an electrical heater could be turned on by low temperature readings from temperature sensors. False temperature readings that are either caused by sensor malfunctions or attacker's malicious modification could trigger the heater staying `on' for long time, which significantly increase the risk of fire.}

\fu{To cope with these security issues, }
there has been many research works~\cite{sikder2019aegis,fu2021hawatcher,choi2018detecting,yamauchi2020anomaly}
\fu{that aim to automatically detect anomalous status of smart home systems. These works model the normal patterns of users' daily activities in the form of invariant event sequences or causal association rules and report deviant patterns as anomalies.} 
For instance, the the event sequence  
\fu{``presence detected'', ``lock unlocked'', ``front door opened'', ``hallway motion active'', ``front door closed'', ``hallway motion inactive'' should be observed when a user goes back to home. Anomaly alarms are raised when the order of this event sequence is violated (e.g., the door gets unlocked without a prior event of presence sensor). }

However, \fu{existing works only focus on the order of events but does not take the events' temporal information into consideration. In some anomalous cases, the sequences of events remains identical to the learned one but with certain events being delayed.}
 
\fu{In the example case of user going back to home}, 
If intruder get the presence sensor and enter the house, the event sequence could also be ``presence detected'', ``lock unlocked'', ``front door opened'', ``hallway motion active'', ``front door closed'', ``hallway motion inactive'', but the time-interval between each two event is different. For instance, while user closes the door immediately after enters the hallway, intruder may leave the door open until he leaves the house. The difference in time-interval is the key to identify anomalous case.

\fu{In this paper, we fill this gap by proposing a novel anomaly detection method that takes events' temporal information into consideration. With the additional information, we are able to more accurately profile normal behaviors of smart home IoT systems. More specifically, we enhance the existing event patterns by adding intervals between each two events in the sequence. In addition, we design an innovative scoring mechanism to compare the temporal similarity between two event sequences. Intervals that are either too small or too large will result in low or high score and trigger anomaly alarms. }

To evaluate the proposed method, we collect the event sequences of user's daily activity \fu{in a real-world testbed over 10 days.} We measure 3 activities that include ``Come back home'', ``Go to work'', ``Use toilet''. The evaluation results show that the accuracy for three activity is 93\%, 88\%, 89\%,  respectively.

 We make the following contributions:
 \begin{itemize}
\item We propose a new learning model that can learning time-interval from event  sequence. With the help of this new learning model, we can better profile user behaviors.
\item We propose a scoring method for event sequence, which gives a score based on the similarity between a test event sequence and the ground truth. The score result is used to decide whether to trigger an anomaly alarm or not. 
\item We evaluate the new learning model and scoring method with a real-world testbed and three activities. The result shows that the accuracy for each activity is 93\%, 88\%, 89\%, respectively.
\end{itemize}

The rest of the paper is organized as follows. We introduce related work in Section~\ref{relatedwork}. In Section~\ref{backg}, we describe background about smart home platform. We introduce motivation and threat model in Section~\ref{motivation}. In Section~\ref{system}, we introduce the design of anomaly detection system. The evaluation is presented in Section~\ref{evaluation}. Conclusion is presented in Section~\ref{conclusion}.

\section{Related work}\label{relatedwork}
With the rapid development of IoT devices and smart home, the security issue in smart home draws much attention and there has been many papers focusing on the anomaly detection in smart home\cite{sikder2019aegis,fu2021hawatcher,choi2018detecting,chi2021pfirewall}. Most of them are based on the order of events. For example, AEGIS observes the change in device behavior based on user activities and builds a contextual model to differentiate benign and malicious behavior\cite{sikder2019aegis}.

Reference \cite{choi2018detecting} precomputes sensor correlation and the transition probability between sensor states and finds a violation of sensor correlation and transition to detect and identify the faults.

The main difference of these existing anomaly detector and our work is that our paper take time-interval into consideration and propose a novel scoring mechanism method for anomaly detection. To the best of our knowledge, time-interval between each two event isn't used for anomaly detection in smart home prior to our work.

\section{Background}\label{backg}
In this section, we \fu{ introduce the background knowledge of} \fu{smart home automation platform and events log.}

\subsection{Smart Homes Automation Platform}

\begin{figure}[htbp]
\centerline{\includegraphics{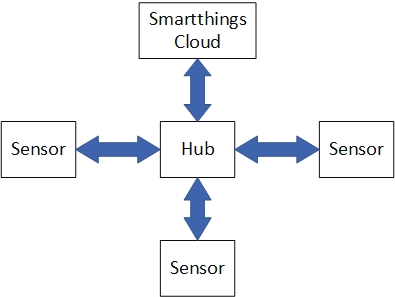}}
\caption{The architecture of smartthings platform}
\label{fig1}
\end{figure}

There are many smart home platforms now, like Smartthings, Amazon Alexa. These smart Home platforms provide so many types of sensor that can be used in smart homes. These sensors have different functions. In this paper, we use smartthings, which is a popular smart platform. 

The architecture of smartthings is shown in Figure~\ref{fig1}. Sensors are connected to each other and smartthings cloud via hub. As a result, sensors can upload its data to smartthings cloud for further analysis and user can control sensor via smarthings cloud.

Automation rule is an important part of smart home platform, which is set by user. When the status of sensor meet the prerequisite of automation rule, the status of other sensor would change. For example, the automation rule could be ``when the door in bathroom open, the light in bathroom should be turned on/off". As a result, when user opens the door to walk into bathroom at the first time, the light in bathroom is turned on. Then when user opens the door to walk out bathroom, the light in bathroom is turned off.

\subsection{Sensor Activity Log}
\begin{table}[h]
\caption{Example events collected from smart home IoT system. }
\begin{center}
\begin{tabular}{|c|c|c|}
\hline
 \textbf{Timestamp} & \textbf{Device} & \textbf{Value} \\ 
 \hline
 10/1/2021 13:00:01& motionSensor & active \\ 
10/1/2021 13:00:02 & light & on \\ 
10/1/2021 13:01:00 & light & off \\ 
......& ...... & ...... \\ 
 \hline
\end{tabular}
\label{sensorlog}
\end{center}
\end{table}
User's acitivity will active multiple sensor, which would be recorded in log. The log can be downloaded from smartthings app, which can be simplified as Table~\ref{sensorlog}. When sensor is activated, the log will record the timestamp, device name and value of the device at this moment. The value could be numerical values, like ``56.0", or boolean value.

\section{\fu{Threat Model}}\label{motivation}

\fu{In this paper, we consider smart home anomalies that are caused by the following reasons:}

\begin{itemize}
\item \textbf{Faulty devices}.  Smart home devices may lost the connection to hub at any time and has no response to user's behavior if they are broken up. When the device is offline, there would be no record about this device in the log. For example, when motion sensor is out of power, it will be offline and lose connection \fu{with the} hub. So, smart home system can't detect any motion and the automation rule based on motion sensor will be invalid.
\item \textbf{Large Delays}. When sensor is activated, it should report the event immediately to hub and this event would be added to log in smartthings cloud. But, due to interference from other devices, sensor may report the event to hub with a large delay. Large delays would cause disorder in log. Also, the automation rule based on the event will be invalid because the prerequisite of it isn't met, which may lead to risky consequence. For example, the door may be left unlocked caused by loss of presence off event when user leaves home\cite{fu2021hawatcher}. 
\item \textbf{Intruder}. We use the event sequence in log to profile user's daily activity. When an intruder enter the building, since its behavior is different from user, the event sequence pattern changes.  For example, when user go back to home, the behavior could be ``open the door, close the door, enter hallway, enter kitchen". But, when intruder enter the building, its behavior may be ``open the door, close the door, enter hallway, enter bedroom ". Since the behavior is different, event sequence will not be the same. Also, even intruder follows the pattern of user's daily behavior, the time-interval between each two event may be not the same. For example, user often takes a nap in bedroom while intruder just takes a look at bedroom, the time interval between ``open the bedroom door" and ``close the bedroom door" would be significantly different.
\item \textbf{Anomalous activities}. Since user's behavior is profiled based on previous collected event sequence. Once anomalous activity happens, the event sequence will be different and can be regarded as anomaly. For example, if user  takes a shower, the event sequence is ``open the door, active motion sensor, close the door, active motion sensor, open the door". But, if user have a fall and get in shock, the event sequence is "open the door, active motion sensor, close the door", which is different from the previous one. 
\end{itemize}

Due to the reasons mentioned above, the anomalous cases of event sequence is different from normal case of event sequence in the following aspects:
\begin{itemize}
\item \textbf{Order of Event}. Compared with the event sequence in normal case, the order of event sequence in anomalous case is different .
\item \textbf{Time-interval}. Time-interval between two adjacent event become larger or smaller in anomalous case.
\end{itemize}

\section{\fu{System Design}}\label{system}
\fu{To cope with the aforementioned anomalies, researchers proposed a series of anomaly detection methods~\cite{jakkula2008anomaly,bakar2016activity} that aim to provide early warnings on anomalous situations before causing damages. Our method improves the existing ones by merging time-intervals between events into the anomaly detection model. In this section, we first introduce our formal representations of event sequence patterns and then discuss how we deal with the additional temporal information using our innovative scoring mechanism.} 

\subsection{\fu{Sequence Pattern Representation}}

\fu{We use $E$ to represent device events. For a specific event of a device, we represent it in the form of $E^{state}_{attr} (Device)$. For example, the event of ``bedroom motion sensor detect active motion'' is represented as  $E^{active}_{motion} (bedroom\ motion)$. Users' common daily activities usually trigger multiple devices' events consecutively, which forms certain sequence patterns. For example, if user use toilet, he would open the door, turn on light and close the door. As a result, contact sensor in the door, light in toilet, motion sensor in toilet are activated. Aside from the sequence of events, we also take the time-interval between each two adjacent events into consideration. In the example above, we also record the interval between ``Open the door, active motion sensor, turn on light, close the door''. As a result, the event sequence pattern of a user's activity can be represented in (\ref{equ:represent})}.

\begin{equation}\label{equ:represent}
    P(Act_{i}) = \left\{
        \begin{aligned}
            &[E^{\alpha}_{A_1} (D_1), E^{\beta}_{A_2}(D_2),\dots,  E^{\gamma}_{A_n}(D_n)] \\
            &[\Delta_1, \Delta_2,  \dots, \Delta_{n-1}]
        \end{aligned}
    \right.
\end{equation}

\fu{ $E^{\alpha}_{A_1} (D_1)$ means the event of attribute $A_1$ on device $D_1$ turns to state $\alpha$. While, $\Delta_i$ stands for the time interval between the $(i)$th and the $(i+1)$th events. }

\subsection{\fu{Scoring Mechanism}}

With the repeat of user's daily activities, there would be many frequent sequences in the log of event sequence. For example, if user use toilet, he would like to open the door, turn on the light and close the door. So after the repeat of this activity,  a frequent event sequence ``open the door, turn on light, close the door" in log is observed. After getting the frequent event sequence, we can extract all event sequence which is identical the frequent event sequence from the log. The we can get the average time-interval of each two event using the extracted event sequence. The frequent event sequence and the vector of average time-interval of each two event can be used to represent the ground truth for this frequent event sequence pattern, which is:
\begin{equation}\label{equ2}
    P(Act_{g}) = \left\{
        \begin{aligned}
            &[E^{\alpha}_{A_1} (D_1), E^{\beta}_{A_2}(D_2),\dots,  E^{\gamma}_{A_n}(D_n)] \\
            &[\overline{\Delta_1}  ,\overline{\Delta_2}  ,  \dots, \overline{\Delta_{n-1} }  ]
        \end{aligned}
    \right.
\end{equation}
Where $E^{\alpha}_{A_1} (D_1), E^{\beta}_{A_2}(D_2),\dots,  E^{\gamma}_{A_n}(D_n)$ is the frequent event sequence, $\overline{\Delta_{i}}$ is the average time interval between event $E^{\beta}_{A_i}(D_i)$ and event $E^{\eta }_{A_{i+1}}(D_{i+1})$.

Compared with the ground truth of event sequence, test event sequence is missing some event or the time-interval between each two event is abnormal. Test event sequence can be represented as:
\begin{equation}\label{equ3}
    P(Act_{t}) = \left\{
        \begin{aligned}
            &[E^{\alpha}_{A_1} (D_1), E^{\beta}_{A_2}(D_2),\dots,  E^{\gamma}_{A_k}(D_k)]\\
            &[\overline{\Delta_1}  ,\overline{\Delta_2}  ,  \dots, \overline{\Delta_{k-1} }  ] \\
        \end{aligned}
    \right.
\end{equation}
Where $P(Act_{t})$ is the test event sequence, k is the number of event and $k\le n$.

We propose a scoring mechanism to quantify the similarity between test event sequence and ground truth, which can be represented as follows:
\begin{equation}
\begin{split}
score&=score_{n}+\alpha *score_{c}  \\
score_{n}&=\frac{num_{t} }{num_{g} } \\
score_{c} &=1-\frac{\theta (ti_{t} ,ti_{g^{'} } )}{\pi } \\
\theta (ti_{t} ,ti_{g^{'} })&\in \left [ 0,\pi  \right ]
\end{split}
\label{score}
\end{equation}
Where $score$ is the score of test event sequence, $num_{t}$ is the number of event in test event sequence, $num_{g}$ is the number of event in ground truth, $\alpha$ is a  coefficient, $ \theta (ti_{t} ,ti_{g^{'} } )$ is the angle between vector $ti_{t} $ and vector $ti_{g^{'}}$, $ti_{t}$ is the time-interval vector of test event sequence, $ti_{g^{'}}$ is the modified time-interval vector of ground truth. Since some event is missing in test event sequence, the event doesn't exist in test event  sequence is deleted from $ti_{g }$ to make sure $ti_{t}$ has the same length with $ti_{g^{'} }$.

\section{Performance Evaluation}\label{evaluation}

\subsection{Experiment Setup}
\begin{figure}[htbp]
\centerline{\includegraphics[scale=0.6]{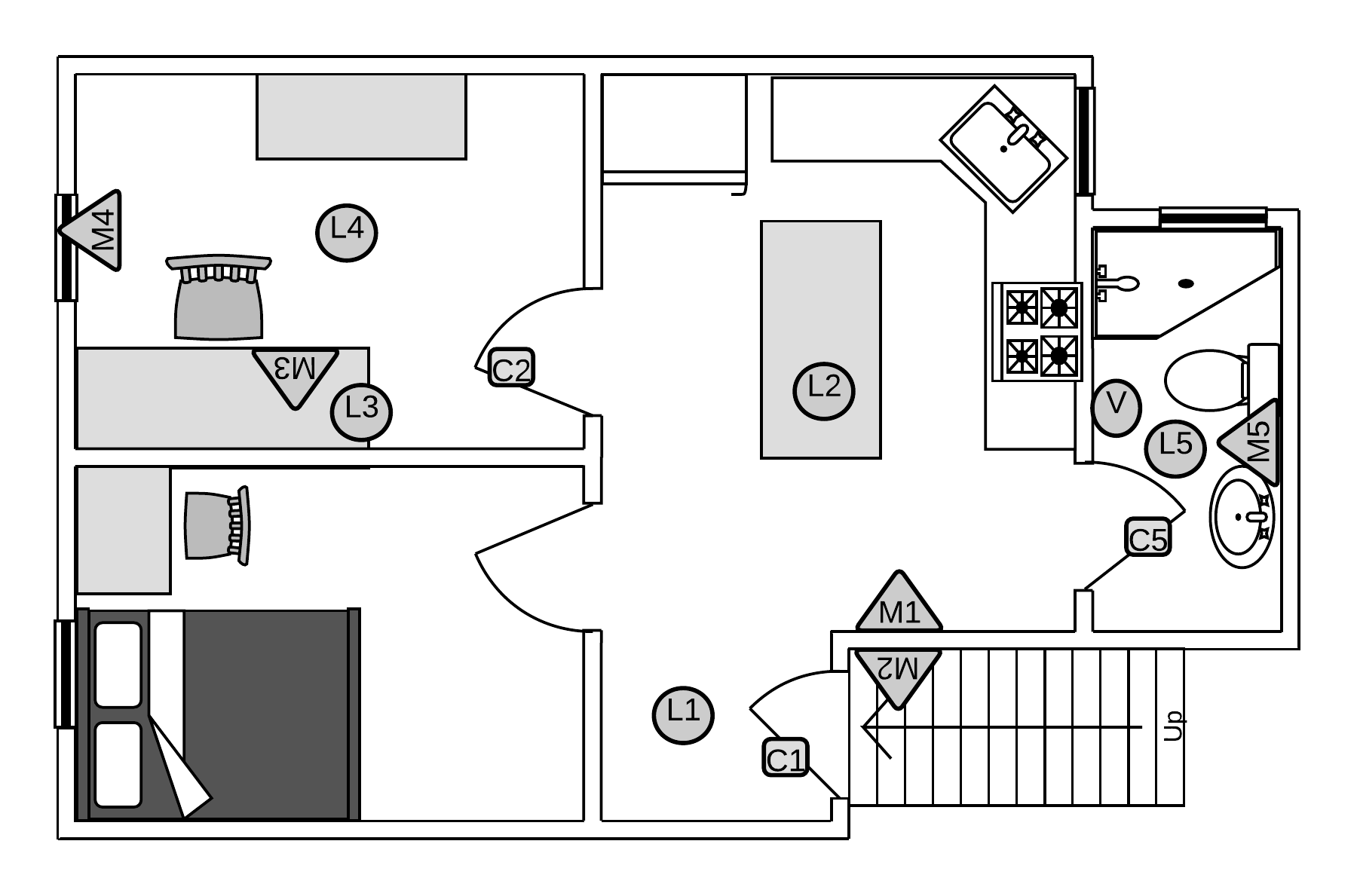}}
\caption{Floor plans of the testbed and device deployment layout.}
\label{testbed}
\end{figure}

We evaluate \fu{our proposed anomaly detection method on a real-world testbed as shown in Figure~\ref{testbed}. All used devices' labels, attributes and installation locations are listed in Table~\ref{device}. The testbed  consists of two bedroom, a dining room, and a bathroom and has one resident. In four rooms, we install 14 IoT devices of 4 types. All devices are connected to a SmartThings hub and managed using the SmartThings mobile app~\cite{smartthings}}. 

\begin{table}[ht]
\caption{IoT devices used in the testbed, their abbreviation labels, attributes and deployment information}
\begin{center}
\begin{tabular}{|c|c|c|c|}
\hline
 \textbf{Label} & \textbf{Device Name}& \textbf{Attributes}&\textbf{Deployment} \\ 
 \hline
 M& \makecell[c]{SmartThings\\ Motion Sensor}&motion&on wall \\ 
 \hline
 C& \makecell[c]{SmartThings\\ Contact Sensor}&contact, acceleration&on doors \\ 
 \hline
 L& \makecell[c]{SmartThings\\ Light Bulb}&switch&as ceiling light, lamp \\ 
 \hline
 V& \makecell[c]{ThreeReality\\ Smart Switch}&switch&to control fan \\ 
 \hline
\end{tabular}
\label{device}
\end{center}
\end{table}

\subsection{\fu{Data Collection and Augmentation}}
\fu{On this testbed, we evaluate our method on 3 activities that include: ``Come back home'', ``Go to work'' and ``Use toilet''. For each activity as listed in Table~\ref{activity}, we deploy some automation rules according to the participant's requirements and collect about 50 instances for each activity during 10 days experiment. All events are collected from SmartThings mobile app.
}

\begin{table}[t]
\caption{Activities used to evaluate the proposed method}
\begin{center}
\begin{tabular}{|c|c|c|c|}
\hline
 \textbf{Activity} & \textbf{Behavior}& \textbf{devices}&

 \textbf{Automation rule} \\ 
 \hline
 \makecell[c]{Come\\ back \\home} & \makecell[l]{User opens and \\closes the front \\door, enters\\ kitchen}& \makecell[c]{M2,\\C1,\\L1,\\M1,\\L2} & \makecell[l]{$E^{active}_{motion}(M2) \rightarrow  E^{on}_{light}(L1)$\\$E^{active}_{motion}(M1) \rightarrow  E^{on}_{light}(L2)$} \\
  \hline
 \makecell[c]{Use\\toilet}&\makecell[l]{User enters the\\ bathroom, closes\\ the door,\\takes a pee,\\opens the door\\ and walks out}&\makecell[c]{M5,\\C5,\\L5,\\V}&\makecell[l]{$E^{active}_{motion}(M5) \rightarrow  E^{on}_{light}(L5)$\\$E^{close}_{contact}(c5) \rightarrow  E^{on}_{switch}(v)$ \\$E^{open>20s}_{contact}(c5)\And$\\ $E^{inactive>20s}_{motion}(M5)\rightarrow $\\$E^{off}_{light}(L5),E^{off}_{switch}(v)$}\\
 \hline
 \makecell[c]{Go\\to\\work}&\makecell[l]{User opens the\\ door, walks\\into study and\\sits before\\the desk}& \makecell[c]{C2,\\M4,\\L4,\\M3,\\L3} &\makecell[l]{$E^{open}_{contact}(c2) \rightarrow  E^{on}_{light}(L4)$\\$E^{active}_{motion}(M3) \rightarrow  E^{on}_{light}(L3)$}\\
 \hline
\end{tabular}
\label{activity}
\end{center}
\end{table}

\fu{To improve the effect of the collected data, we augment the collected events of activity instances. More specifically, }we use synthetic minority over-sampling technique (SMOTE)~\cite{chawla2002smote} to expand the collected dataset.  SMOTE is mainly used in over-sampling the minority (abnormal) class and under-sampling the majority (normal) class to achieve better classifier performance. 
\fu{Because the difference between nearest neighbor and the selected data would be tiny, we randomly select two instances of normal data rather than select k-nearest  neighbors.}
The detailed process can be represented as:
\begin{equation}
\begin{split}
P_{new} =P_{i}+0.5*(P_{j}-P_{i}) 
\end{split}
\end{equation}
Where $P_{i}$ and $P_{j}$ is the vector of the collected data, $P_{new}$ is the vector of the created data.

\fu{Since anomalies rarely happen during the experiment period, we generate data of abnormal instances by simulating anomalies based on data of normal instances.} 
\fu{We simulate anomaly cases in two methods: 1) deleting one event from a normal event sequence; (2) extend one time interval in an event sequence by 50 times. We call these two types of anomaly cases as ``Anomaly(seq)'' and ``Anomaly(ti)'', respectively.}

\begin{figure}[htbp]
\centerline{\includegraphics[scale=0.6]{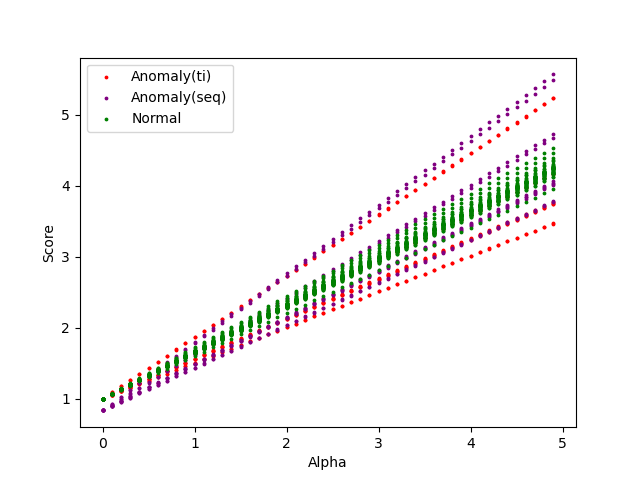}}
\caption{The influence of alpha on score. Green points are scores of normal data instances, purple points are for anomaly(seq), red point are for anomaly(ti).}
\label{scorealpha}
\end{figure}

\subsection{\fu{Training Process}}

In the training process, we use collected normal and anomaly instances to find the optimal $\alpha$ and score section. We use 40 instances of normal data, 10 instances of anomaly(seq) data, and 10 instances of anomaly(ti) for each activity.

Equation~(\ref{score}) shows that we can calculate the score of test event sequence by comparing it with a given reference sequence. The score is affected by the value of $\alpha$.
To explore the impact of the value of $\alpha$, we calculate the score of all data instances in the testing dataset with different $\alpha$ values in the range of $[0,5]$.    
From Figure~\ref{scorealpha}, we can select the $\alpha$ that the cluster of green point is obviously divorce from the cluster of purple point and red point. For each selected $\alpha$, we can find the score section in which green point dominate over purple point and red point. So this score section can be selected as the threshold to distinguish normal data instances with anomaly instances.

By setting $\alpha$ as $3$, we can get the score threshold boundaries of $[2.9, 3.1]$, $[3.45, 4]$, and $[2.5, 2.62]$ for normal activities of "Come back home", "Go to work", and "Use toilet", respectively.

\subsection{Testing Results}
In the testing process, we apply the learned score range in the training process to 100 testing data instances for each activity to check the performance of anomaly detection. Each activity contains 20 instances of anomaly(seq) data, 20 instances of anomaly(ti), and 60 instances of normal data. None of the testing data is used in the training process. 

We use True Positive (TP), False Negative (FN), False Positive (FP), True Negative (TN) and accuracy (represented as the eqution shown below) to evaluate the result on three activities. The testing result for each activity is shown in Table \ref{resulthome}, Table \ref{resultwork} and Table \ref{resulttoilet}.

\begin{equation}
\begin{split}
Accuracy=\frac{TP+TN}{TP+TN+FP+FN} 
\end{split}
\end{equation}

\begin{table}[htbp]
\caption{Testing results of activity ``Come back home''}
\begin{center}
\begin{tabular}{|c|c|c|c|c|}
\hline
 &\textbf{Amount} & \textbf{TP+TN}& \textbf{FN+FP}&\textbf{Accuracy} \\ 
 \hline
Anomaly(seq)& 20 & 20 &0& 100\% \\
Anomaly(ti)& 20 & 16 &4& 80\%  \\ 
Normal& 60 & 57 &3 & 95\%           \\ 
\textbf{Total}& \textbf{100}&\textbf{93}&\textbf{7}&\textbf{93\%}\\
 \hline
\end{tabular}
\label{resulthome}
\end{center}
\end{table}
Table \ref{resulthome} is the testing result of activity ``Come back home''. The average accuracy for all category is 93\%. Accuracy of anomaly(seq), anomaly(ti), normal is 100\%, 80\%, 95\%, respectively. 

\begin{table}[htbp]
\caption{Testing results of activity ``Go to work''}
\begin{center}
\begin{tabular}{|c|c|c|c|c|}
\hline
 &\textbf{Amount} & \textbf{TP+TN}& \textbf{FN+FP}&\textbf{Accuracy} \\ 
 \hline
Anomaly(seq)& 20 & 9 &11& 45\% \\
Anomaly(ti)& 20 & 20 &0& 100\%  \\ 
Normal& 60 & 59 &1 & 95\%           \\ 
\textbf{Total}& \textbf{100}&\textbf{88}&\textbf{12}&\textbf{88\%}\\
 \hline
\end{tabular}
\label{resultwork}
\end{center}
\end{table}
Table \ref{resultwork} is the testing result of activity ``Go to work''. The average accuracy for all category is 88\%. Accuracy of anomaly(seq), anomaly(ti), normal is 45\%, 100\%, 95\%, respectively. 

\begin{table}[htbp]
\caption{Testing results of activity ``Use toilet''}
\begin{center}
\begin{tabular}{|c|c|c|c|c|}
\hline
 &\textbf{Amount} & \textbf{TP+TN}& \textbf{FN+FP}&\textbf{Accuracy} \\ 
 \hline
Anomaly(seq)& 20 & 20 &0& 100\% \\
Anomaly(ti)& 20 & 9 &11& 45\%  \\ 
Normal& 60 & 60 &1 & 100\%           \\ 
\textbf{Total}& \textbf{100}&\textbf{89}&\textbf{11}&\textbf{89\%}\\
 \hline
\end{tabular}
\label{resulttoilet}
\end{center}
\end{table}
Table \ref{resulttoilet} is the testing result of activity ``Use toilet''. The average accuracy for all category is 89\%. Accuracy of anomaly(seq), anomaly(ti), normal is 100\%, 45\%, 100\%, respectively.

\section{Conclusion}\label{conclusion}
In this paper, we proposed a new learning model that takes time interval among events into consideration. We designed a new score metric to quantify the temporal similarity between a testing event sequence and a reference sequence of ground truth. We evaluated the performance of the proposed method with 10 days experiment in a real-world testbed. The results showed our method achieves the accuracies of 93\%, 88\%, and 89\% for three testing daily activities.

\bibliography{refs} 
\end{document}